\documentclass[letterpaper]{article}
\usepackage{aaai}
\usepackage{times}
\usepackage{helvet}
\usepackage{courier}
\usepackage{graphicx}
\usepackage{amsmath}
\usepackage{float}
\usepackage{amsfonts}
\usepackage{amssymb}
\nocopyright
\usepackage{url}  %Required
\usepackage{graphicx}  %Required

\begin{document}
\title{Knowledge Tracing Challenge:\\
Optimal Activity Sequencing for Students}
\author{Yann Hicke  \\
  Cornell University\\
  Ithaca, USA \\
  \texttt{ylh8@cornell.edu}}
\maketitle
\begin{abstract}
\begin{quote}
Knowledge tracing is a method used in education to assess and track the acquisition of knowledge by individual learners. It involves using a variety of techniques, such as quizzes, tests, and other forms of assessment, to determine what a learner knows and doesn't know about a particular subject. The goal of knowledge tracing is to identify gaps in understanding and provide targeted instruction to help learners improve their understanding and retention of material. This can be particularly useful in situations where learners are working at their own pace, such as in online learning environments. By providing regular feedback and adjusting instruction based on individual needs, knowledge tracing can help learners make more efficient progress and achieve better outcomes. Effectively solving the KT problem would unlock the potential of computer-aided education applications such as intelligent tutoring systems, curriculum learning, and learning materials’ recommendation. 
In this paper, we will present the results of the implementation of two Knowledge Tracing algorithms on a newly released dataset as part of the AAAI2023 Global Knowledge Tracing Challenge.
\end{quote}
\end{abstract}

\section{Introduction}
Knowledge tracing (KT) is an approach to educational research and assessment that focuses on modeling how learners acquire and retain knowledge. The goal of KT is to understand how learners interact with educational materials and to identify the most effective instructional strategies to improve learning outcomes. \newline
Knowledge tracing has its roots in the work of cognitive psychologists, such as Anderson who used mathematical models to examine how humans acquire and store information. Specifically, Anderson proposed a model of learning called the Adaptive Control of Thought (ACT) theory \cite{anderson1993problem}, which posits that humans construct mental representations of the world around them, and that these representations are used to guide behavior. This theory provided a foundational framework for understanding how learners interact with educational materials.\newline
In the 1990s, researchers began to apply the principles of ACT to educational settings. They developed a variety of computer-based KT systems, such as Bayesian Knowledge Tracing (BKT) \cite{corbett1994knowledge} which used machine learning algorithms to track student learning. These systems provided researchers with an empirical approach to understanding student learning, and they proved to be useful tools for assessing the effectiveness of instructional interventions.
\newline
The theoretical foundations of KT are grounded in the cognitive science literature: the ACT theory of learning. In this framework, learners are viewed as active agents who are constantly constructing new knowledge based on their experiences. In addition, KT is informed by the idea of constructivism, which suggests that learners construct their own understanding of the world based on their experiences. Knowledge tracing has been applied to a variety of educational settings, including K-12 classrooms, higher education courses, and online learning platforms. In each of these contexts, KT has been used as a tool for assessing student learning and identifying effective instructional strategies. By tracking student performance on assessments, researchers can gain insights into which instructional strategies are most effective for different types of learners. Despite its potential, there are still some challenges associated with KT, such as the lack of standardized methods for data collection and analysis. Nonetheless, KT has the potential to revolutionize educational research and assessment, and its use is likely to continue to grow in the coming years. \\
\begin{figure*}[h!]
    \centering    \includegraphics[width=0.9\textwidth]{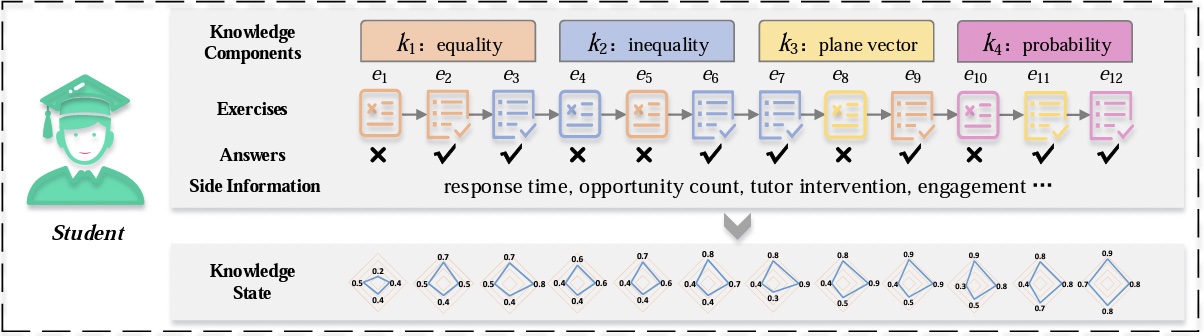}
    \caption{Schematic diagram of Knowledge Tracing\protect\footnotemark}
    \label{fig:explanation_KT}
\end{figure*}
\footnotetext{Image created by \cite{liu2021survey}}
\newline
To better understand the KT problem, an example is usually helpful. In Figure 1, a sequence of events is depicted. It shows an interaction scenario between a student and an Intelligent Tutoring System (ITS) in which the student is given a sequence of exercises taken from an exercise set $\{e_1,\cdots,e_{12}\}$ and asked to solve these exercises. During the interaction, the ITS estimates the student’s knowledge states over the knowledge components $\{k_1, k_2,k_3,k_4\}$ (here: equality, inequality, plane vector, probability) that are required to solve these exercises. A knowledge component is defined as a description of a mental structure or process that a learner uses, alone or in combination with other knowledge components, to accomplish steps in a task or a problem \cite{koedinger2012knowledge}. However, capturing a student’s knowledge state is a challenging task due to several reasons: 
\begin{itemize}
    \item Each exercise might require more than one knowledge component, which adds complexity to trace knowledge states.
    \item Dependency among skills is another important factor to consider when tackling the KT problem. For example, although $k_3$ requires only skill $k_1$, $k_1$ and $k_2$ are prerequisites for $k_4$ according to the dependency graph shown in Figure \ref{fig:Dependency_graph}.
    \item A student’s forgetting behavior \cite{ebbinghaus2013memory} may result in decaying their knowledge over skills. By modeling forgetting features, skills can be ranked by their relevance to forgetting. For example, Figure \ref{fig:Forgetting_rise} shows that skill $k_4$ is least affected by forgetting when the latest exercise $e_{12}$ is reached, whereas skill $k_1$ is the most affected one.
\end{itemize}
\begin{figure}[h!]
    \centering
\includegraphics[width=0.2\textwidth]{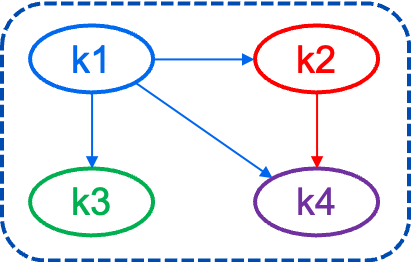}
    \caption{Dependency Graph between knowledge components}
    \label{fig:Dependency_graph}
\end{figure}

\begin{figure}[h!]
    \centering
\includegraphics[width=0.2\textwidth]{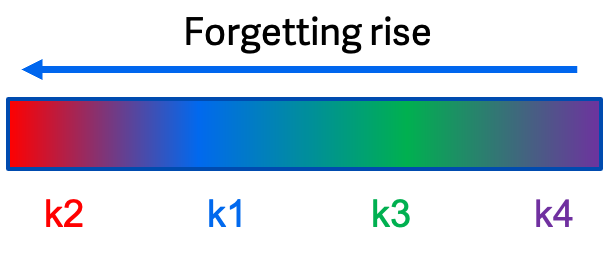}
    \caption{Forgetting of the knowledge components}
    \label{fig:Forgetting_rise}
\end{figure}
The aim of this research is to implement two KT algorithms and evaluate their accuracy on a newly released dataset as part of the AAAI2023 Global Knowledge Tracing Challenge.\newline
The research questions that are posed are the following:
\begin{itemize}
    \item RQ1: What KT algorithm will perform best?
    \item RQ2: Why does the previously found best algorithm perform better?
\end{itemize}

\section{Related work}

\begin{figure}[H]
    \centering
    \includegraphics[scale=0.4]{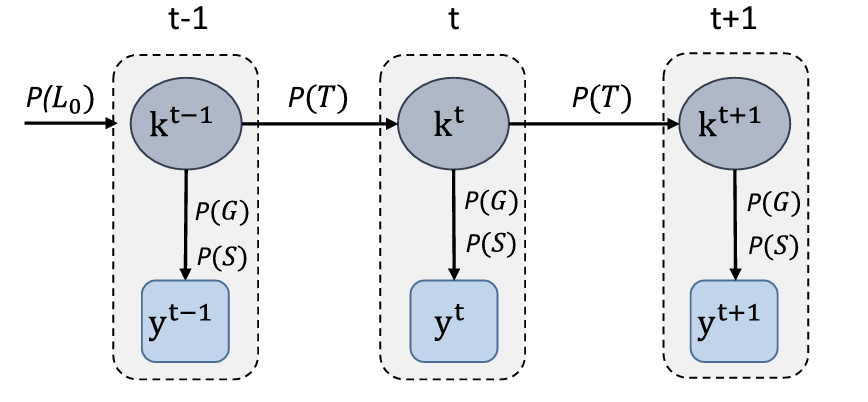}
    \caption[Caption for LOF]{Bayesian Knowledge Tracing\protect\footnotemark}
    \label{fig:BKT}
\end{figure}

Knowledge tracing (KT) is a well-studied topic in educational technology, particularly within the field of Intelligent Tutoring Systems (ITS).
\newline
\newline
\textbf{Probabilistic Models}
\newline
Early work in KT focused on the use of Bayesian networks for modeling student knowledge \cite{anderson1995cognitive}. These models were used to predict student performance and identify areas of difficulty. BKT is a special case of Hidden Markov Model (HMM). There are two types of parameters in HMM: transition probabilities and emission probabilities. In BKT, the transition probabilities are defined by two learning parameters: (1) P(T), the probability of transition from the unlearned state to the learned state; (2) P(F), the probability of forgetting a previously known KC, which is assumed to be zero in BKT. And for the emission probabilities two performance parameters are used: (1) P(G), the probability that a student will guess correctly in spite of non-mastery; (2) P(S), the probability a student will make a mistake (probably of the student to "slip") in spite of mastery. Furthermore, the parameter P($L_0$) represents the initial probability of mastery. This Hidden Markov Model is depicted in \ref{fig:BKT}.
\footnotetext{Image created by \cite{abdelrahman2022knowledge}}

\begin{figure*}[h!]
    \centering
    \includegraphics[width=0.6\textwidth]{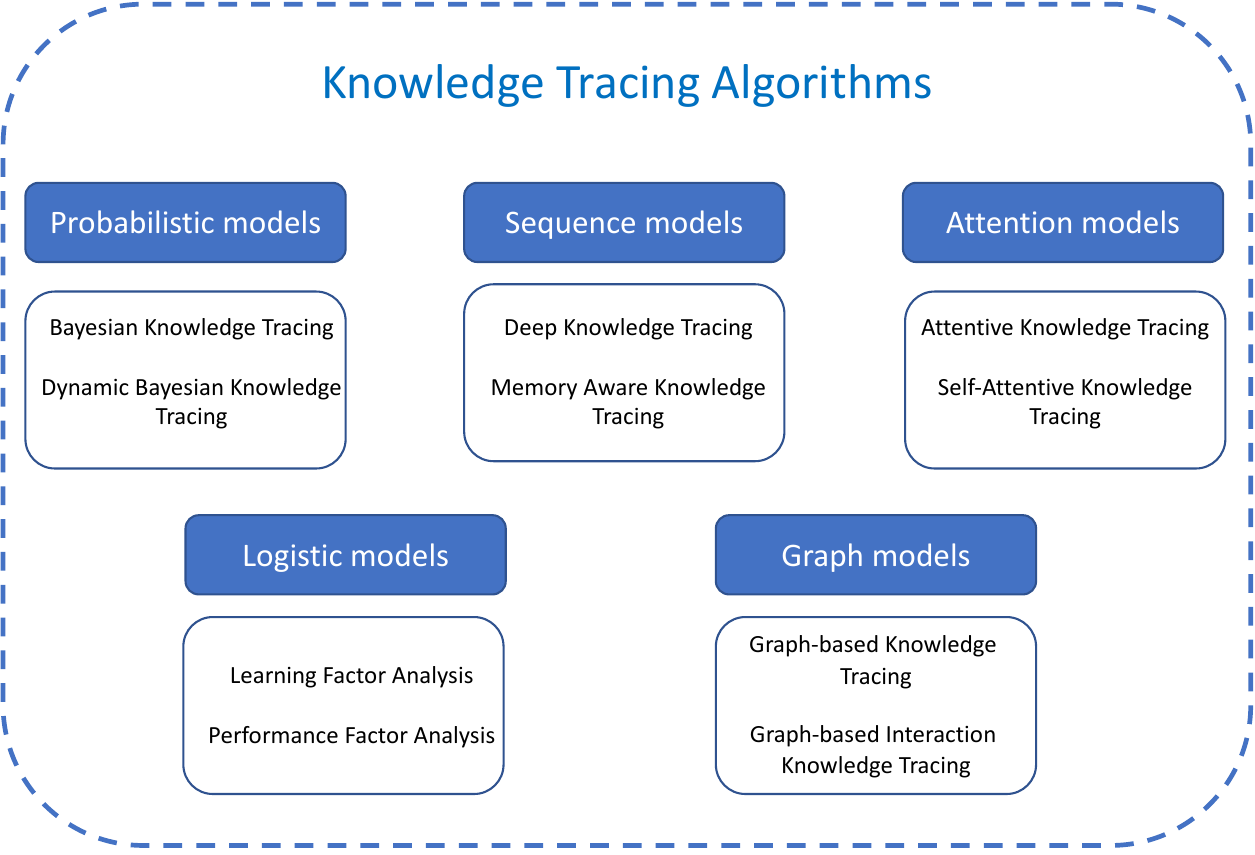}
    \caption{Knowledge Tracing algorithms}
    \label{fig:KT_algos}
\end{figure*}

There are many other models that attempt to solve the KT problem. This paper is not a literature review yet a broad summary is included in Figure \ref{fig:KT_algos}.

\section{Methods}
\subsection{Data \& Task Description}
The data that is studied is a large student assessment dataset with rich textual and structural information (well anonymized). The data is collected from students in grade 3 math classes at Xueersi from TAL Education Group which is a leading smart learning solutions provider in China. Specifically, the dataset is made up of 18,066 students, 7,652 questions, 1,175 KCs, and 5,549,635 interactions/responses. The average student historical sequence length is 307.19. 79.47\% responses are with positive labels, i.e., students correctly answer the questions.
\newline
Each question is associated with question text content, the explanations of question answers and its corresponding KCs. The relevant KCs are hierarchically structured. All the textual characters are anonymized and mapped into internal unique integer tokens.

\subsection{Evaluation Task}
There are 3,613 student interaction sequences in the test set. For each interaction sequence, the first 50\% of interactions are known and the task is to predict the rest 50\% interactions’ responses, i.e., 0 or 1.
\newline
We will choose to use AUC score as the main evaluation metric for this competition.

\subsection{Data Files}
\begin{figure}[H]
    \centering
\includegraphics[scale=0.5]{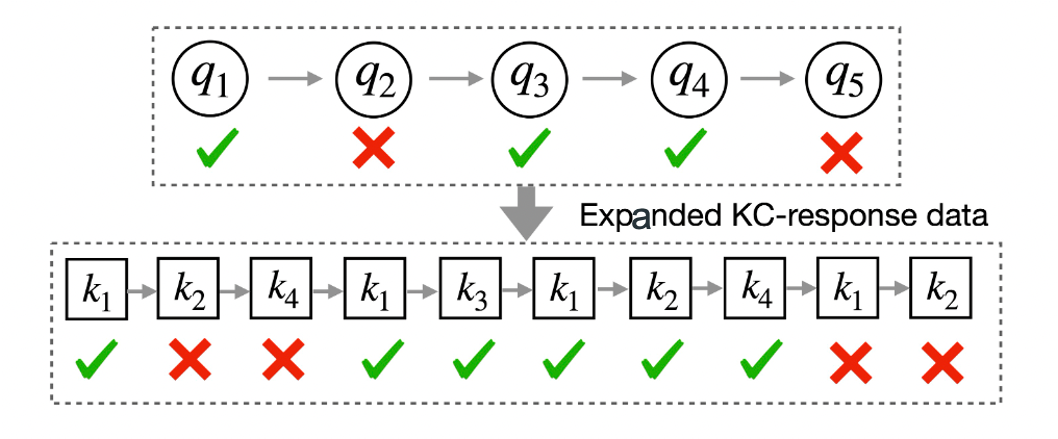}
    \caption{Expanded questions}
    \label{fig:expanded_questions}
\end{figure}

\textbf{train\_valid\_sequences.csv:} the main data file to conduct offline model training and validation. Each student interaction sequence at question level is first expanded into KC level when a question is associated with multiple KCs as shown in Figure \ref{fig:expanded_questions}.
Then each sequence is truncated into sub-sequences of length of 200. The data has been split into 5 folds. The file has the following columns:
\begin{itemize}
    \item fold: the id for each fold, ranging from 0, 1, 2, 3, 4.
    \item uid: internal user id.
questions: the sequence of internal question ids.
    \item concepts: the corresponding sequence of internal KC ids.
    \item timestamps: regular timestamps.
    \item selectmasks: if the sub-sequence is less than 200, we add paddings to make it to 200. “-1” indicates these paddings.
    \item is\_repeat: “1” indicates that the current KC and its previous KC belong to the same question.
\end{itemize}
\textbf{pykt\_test.csv:} the main file for performance evaluation. Each row in pykt\_test.csv represents a test student interaction sequence. There are 3,613 student interaction sequences in the test set. For each interaction sequence, the first 50\% of interactions are known and the task is to predict the rest 50\% interactions’ responses, i.e., 0 or 1. The responses marked as “-1” are those responses to be predicted. Columns of uid, questions, concepts, responses, timestamps, is\_repeat are consistent with columns in train\_valid\_sequences.csv.
\newline
\newline
\textbf{keyid2idx.json:} the id mapping file that stores the id mappings between the original question/KC/user ids and the internal question/KC/user ids. For example, "questions": {"355": 0, "1545": 1, "827": 2, "316": 3, "1638": 4, ..., 355 is the original question id and its corresponding internal question id is 0.
\newline
\newline
\textbf{questions.json:} the file that contains all the auxiliary information of questions and KCs. The json file structure is defined as follows:
\begin{figure}[H]
    \centering
    \includegraphics[scale=0.5]{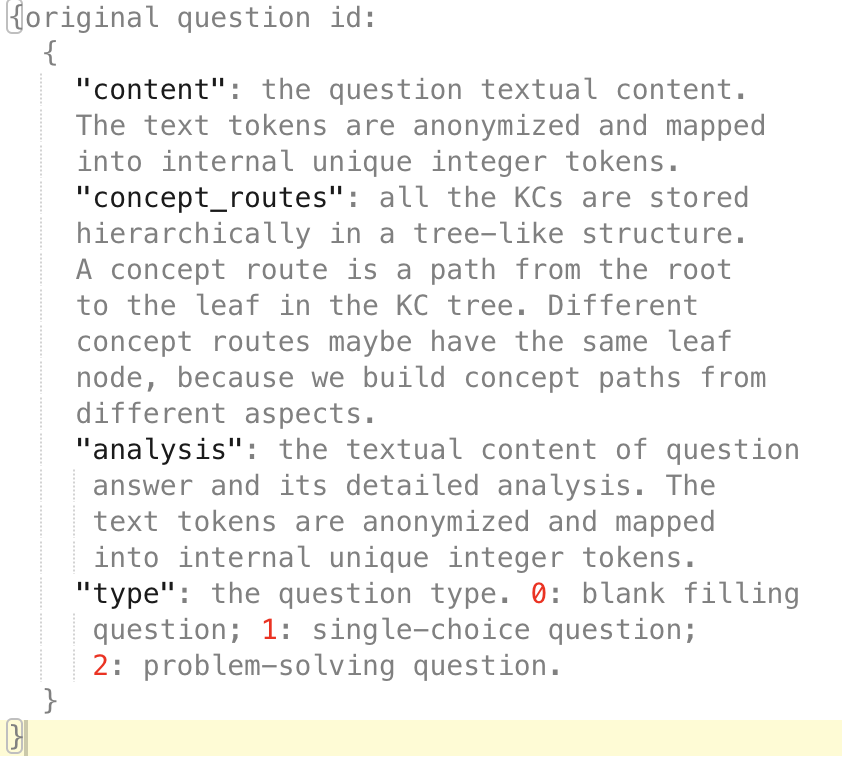}
    \caption{questions.json}
    \label{fig:questions_json}
\end{figure}

\subsection{Implemented algorithms}
We chose to implement two algorithms on this new dataset: DKT and AKT. DKT (Deep Knowledge Tracing) - an RNN-based model - is a reference in Knowledge Tracing; it has been standing at the top of the most effective algorithms for more than five years. It is a simple yet effective algorithm which we are describing below.
\newline
The second algorithm that we are implementing is AKT (Context-Aware Attentive Knowledge Tracing) - a transformer-based model which has now taken over as the front-runner on most KT datasets. 
\newline
\newline
\noindent
\textbf{DKT:} Deep learning methods have been applied to KT, with neural networks such as Recurrent Neural Networks (RNNs), and quickly after Long Short-Term Memory (LSTM) being used to capture temporal dynamics in student learning \cite{piech2015deep}.
\newline
Let's define a set of exercises $\mathcal{E} = \{e_1, \cdots, e_n\}$.
When a student interacts with the exercises in $\mathcal{E}$, a sequence of interactions $\mathcal{X} = ⟨x_1,x_2,\cdots,x_{t-1}⟩$ undertaken by the student can be observed, where $x_i$ = ($e_i$,$y_i$) consisting of an exercise $e_i$ and answer $y_i \in \{0,1\}$. $y_i$ = 0 means that $e_i$ is incorrectly answered and $y_i$ = 1 means that $e_i$ is correctly answered.
\newline
Given a sequence of interactions $X$ that contains the previous question answering of a student, the knowledge tracing problem is to predict the probability $p_t$ of correctly answering a new question $q_t$ at the time step $t$ by the student, i.e., $p_t=\left(y_t=1 \mid q_t, X\right)$. A sequence of hidden states ⟨$h_1, h_2, \cdots, h_n$⟩ is computed which encodes the sequence information obtained from previous interactions. At each time step $t$, the model calculates the hidden state $h_t$ and the student's response $p_t$ as follows:
\begin{align}
h_t=\operatorname{Tanh}\left(W_{h x} x_t+W_{h h} h_{t-1}+b_h\right) \\
p_t=\sigma\left(W_{h y} h_t+b_p\right)
\end{align}
\newline
We then just convert $p_t$ into a 1 or 0 based on whether is greater or less than 0.5.
\newline
After an extensive amount of pre-processing it only requires us to apply an LSTM architecture to our problem.
\newline
\newline
\textbf{Attentive Knowledge Tracing:} The second architecture that we implemented is a transformer-based architecture.
 \cite{ghosh2020context}  presented a contextaware attentive knowledge tracing (AKT) model, incorporating self-attention mechanism with cognitive and psychometric models. AKT comprises four modules: Rasch model based embeddings, exercise encoder, knowledge encoder and knowledge retriever. These modules are depicted on Figure \ref{fig:akt}.
\begin{figure*}
    \centering
\includegraphics[scale=0.47]{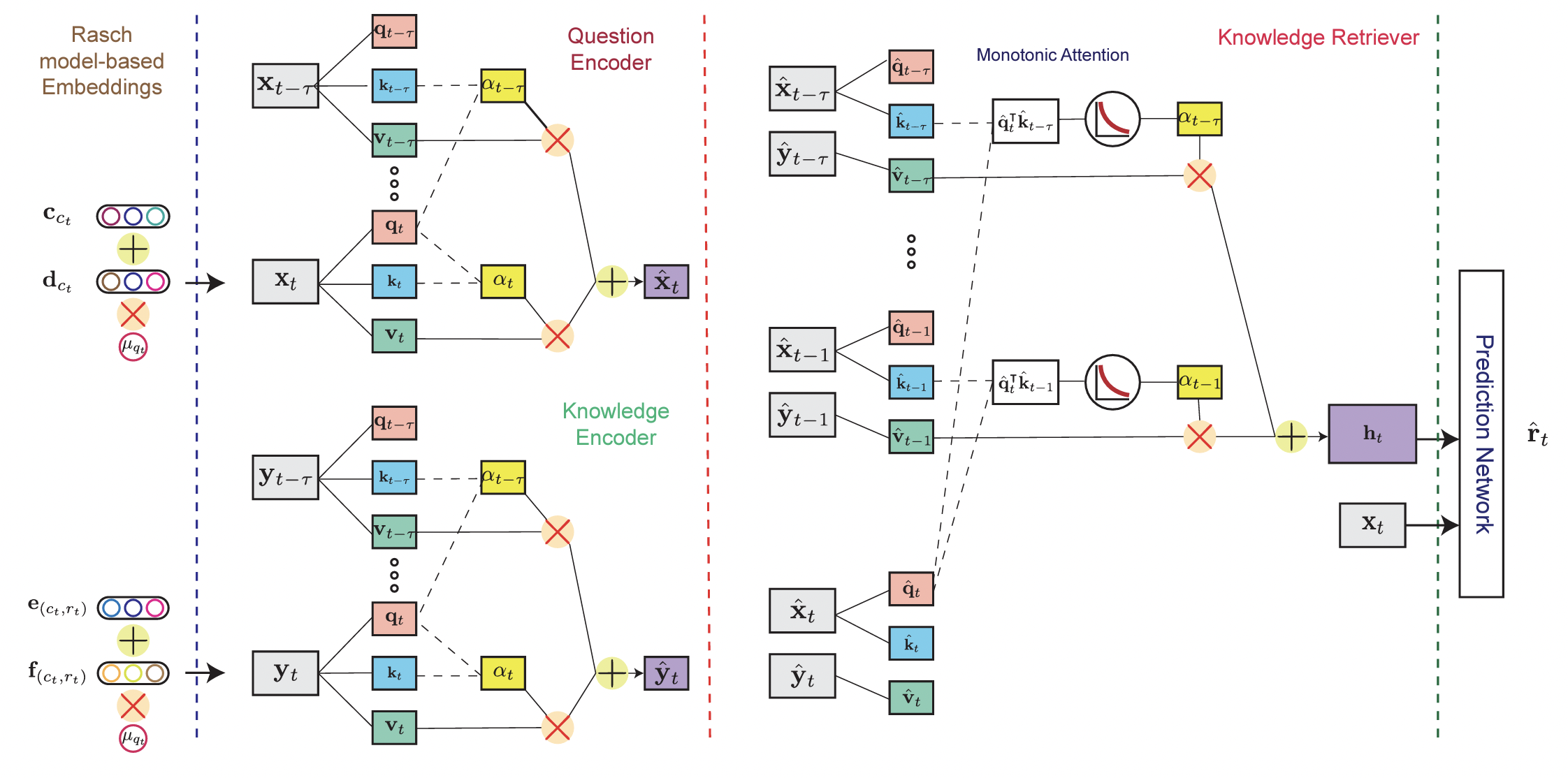}
    \caption{AKT model\protect\footnotemark}
    \label{fig:akt}
\end{figure*}
\footnotetext{Image created by \cite{ghosh2020context}}
 More specifically, the classic and powerful Rasch model in psychometrics \cite{lord2012applications} was utilized to construct embeddings for exercises and KCs. The embedding of the exercise $e_t$ with KC $c_t$ is constructed as follows:
\newline
$$
\boldsymbol{x}_t=\boldsymbol{c}_{c_t}+\mu_{e_t} \cdot \boldsymbol{d}_{c_t}
$$
where $\boldsymbol{c}_{c_t} \in \mathbb{R}^D$ is the embedding of the $\mathrm{KC}$ of this exercise, $\boldsymbol{d}_{c_t} \in \mathbb{R}^{\boldsymbol{D}}$ is a vector that summarizes the variation in exercises with the related $\mathrm{KC}$, and $\mu_{e_t} \in \mathbb{R}^D$ is a scalar difficulty parameter that controls the extent to which this exercise deviates from the related $\mathrm{KC}$. The exercise-answer tuple $\left(e_t, a_t\right)$ is similarly extended using the scalar difficulty parameter for each pair:
$$
\boldsymbol{y}_t=\boldsymbol{q}_{\left(c_t, a_t\right)}+\mu_{e_t} \cdot \boldsymbol{f}_{\left(c_t, a_t\right)},
$$
where $\boldsymbol{q}_{\left(c_t, a_t\right)} \in \mathbb{R}^{\boldsymbol{D}}$ and $\boldsymbol{f}_{\left(c_t, a_t\right)} \in \mathbb{R}^{\boldsymbol{D}}$ are KC-answer embedding and variation vectors. Through such embedding, exercises labeled as the same $\mathrm{KC}$ s are determined to be closely related while keeping important individual characteristics. Then, the input of the exercise encoder is the exercise embeddings $\left\{\boldsymbol{e}_1, \ldots, \boldsymbol{e}_t\right\}$ and the output is a sequence of context-aware exercise embeddings $\left\{\widetilde{\boldsymbol{e}}_1, \ldots, \widetilde{\boldsymbol{e}}_t\right\}$. AKT designs a monotonic attention mechanism to accomplish the above process, where the context-aware embedding of each exercise depends on both itself and the previous exercises, i.e., $\widetilde{\boldsymbol{e}}_t=f_{e^{n c} c_1}\left(\boldsymbol{e}_1, \ldots, \boldsymbol{e}_t\right)$. Similarly, the knowledge encoder takes exercise-answer embeddings $\left\{\boldsymbol{y}_1, \ldots, \boldsymbol{y}_t\right\}$ as input and outputs a sequence of context-aware embeddings of the knowledge acquisitions $\left\{\widetilde{\boldsymbol{y}}_1, \ldots, \widetilde{\boldsymbol{y}}_t\right\}$ using the same monotonic attention mechanism, which are also determined by students' answers to both the current exercise and prior exercises, i.e., $\widetilde{\boldsymbol{y}}_t=f_{e n c_1}\left(\boldsymbol{y}_1, \ldots, \boldsymbol{y}_t\right)$. Finally, the knowledge retriever takes the context-aware exercise embedding $\widetilde{\boldsymbol{e}}_{1: t}$ and exercise-answer pair embeddings $\widetilde{\boldsymbol{y}}_{1: t}$ as input and outputs a retrieved knowledge state $\boldsymbol{h}_t$ for the current exercise. Since the student's current knowledge state depends on the related answering exercise, it is also contextaware in AKT. The novel monotonic attention mechanism proposed in $\mathrm{AKT}$ is based on the assumptions that the learning process is temporal and students' knowledge will decay over time. Therefore, the scaled inner-product attention mechanism utilized in the original transformer is not suitable for the KT task. AKT uses exponential decay and a context-aware relative distance measure to computes the attention weights. Since AKT achieves outstanding performance on predicting students' future answers, as well as demonstrating interpretability due to the combination of cognitive and psychometric models it was our main choice as an algorithm to implement.

\section{Results}
\subsection{Baselines}
DKT and AKT are also two useful algorithms for our experiments since they tend to be baselines that we can compare our implementation against. On the aforementioned training data they were implemented and we display in Figure \ref{fig:baselines} their performance. Practically, there are two different approaches, i.e., \textbf{accumulative prediction} and \textbf{non-accumulative prediction}. The accumulative prediction approach uses the last predicted values for the current prediction while the non-accumulative prediction predicts all future values all at once.
All these approaches are purely trained with question/KC ids and student responses without any auxiliary information, such as question content. The results are used as baseline results for the AAAI2023 competition:
\begin{figure}[H]
    \centering
\includegraphics[scale=0.45]{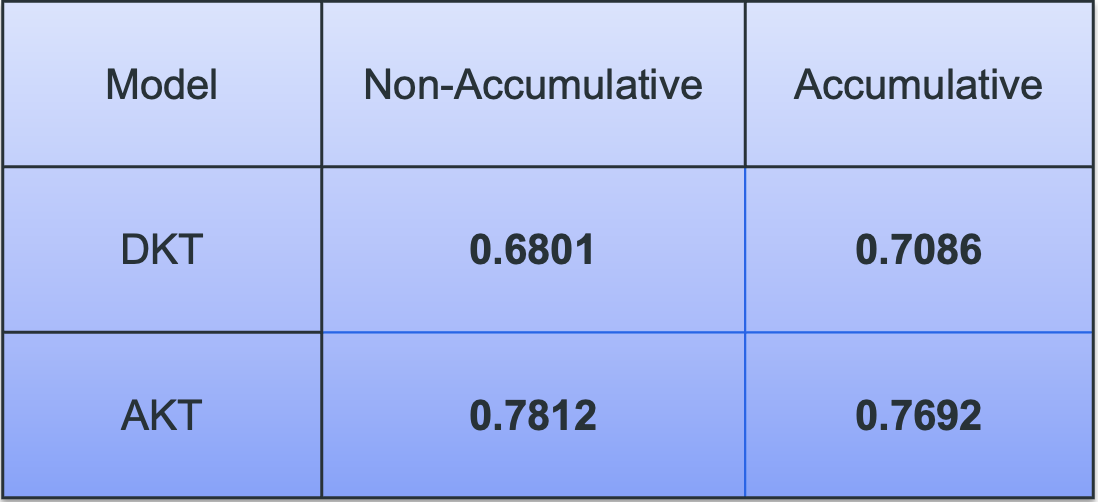}
    \caption{Baselines to beat}
    \label{fig:baselines}
\end{figure}
\subsection{Implementations performance}
Figure \ref{fig:results} shows the performance of our implemented algorithms.
\begin{figure}[H]
    \centering
    \includegraphics[scale=0.4]{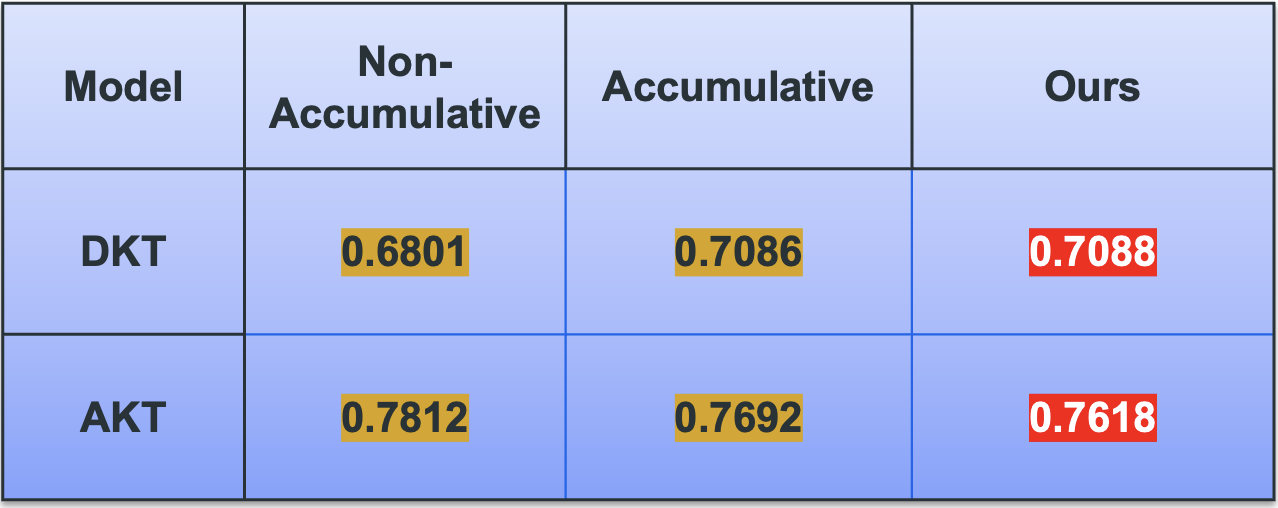}
    \caption{Our results}
    \label{fig:results}
\end{figure}
\noindent
We did manage to perform better than the DKT baseline with our implementation of the Deep Knowledge Tracing algorithms. However, we struggled to reach similar performance on AKT due to a lack of resources for Hyperparameter optimization. One run of our algorithm on our local GPU (M2 chip) takes 28 hours as shown in Figure 
\ref{fig:gpu}.
\begin{figure}[H]
    \centering
    \includegraphics[scale=0.6]{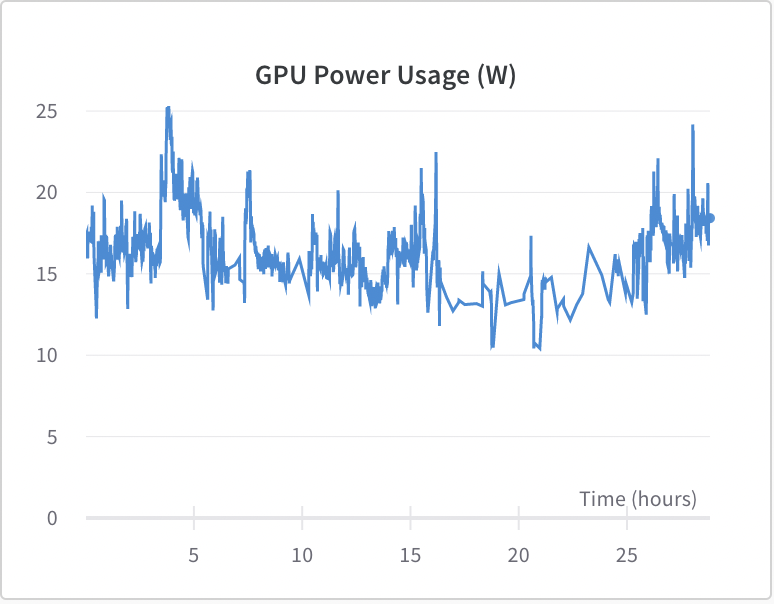}
    \caption{GPU training time}
    \label{fig:gpu}
\end{figure}
For this reason it is challenging to find the right hyperparameters. We did end up using a GCP instance with a Tesla P100 GPU and reduced the training time to 90 minutes. However, it comes at a financial cost.

\section{Discussion}
It is no surprise that AKT would perform better. It uses both psychometrics design principles which are very useful to model student learning and an attention-based architecture well-suited to capture long dependencies like those that we can find in a Knowledge Tracing problem as inspired by long sequence dependencies in NLP. Indeed, Transformers have been pivotal in NLP research on the machine translation task \cite{vaswani2017attention}. This model gets rid of recurrence entirely and uses attention instead to model dependencies inside long sequences. Transformers solved the vanishing and exploding gradients problem which hindered RNNs-based architectures to maintain a good understanding of long dependencies. When introduced in 2017 transformers demonstrated their superior ability in feature extraction while keeping their computational requirement relatively low. The first transformer-based model which obtained State-Of-The-Art results on many NLP tasks is BERT \cite{devlin2018bert}. Following these SOTA results and the general concept of attention \cite{pandey2019self} proposed a self-attentive model for knowledge tracing (SAKT), which directly applied the transformer to capture long-term dependencies between students’ learning interaction.
\newline
However, the complexity of the KT task, which is caused by the interactions between students and exercises, limited the performance of a simple direct transformer application. Therefore, \cite{choi2020towards} proposed a new model named separated self-attentive neural knowledge tracing (SAINT) to improve self-attentive computation for knowledge tracing adaptation. More specifically, SAINT has an encoder-decoder structure, where the exercise and answer embeddings are separately encoded and decoded by self-attention layers. This separation of input allows SAINT to stack self-attention layers multiple times and capture complex relations among exercises and answers. Subsequently, the SAINT+ model \cite{shin2021saint+} was proposed to incorporate two temporal features into SAINT: namely, the answering time for each exercise and the interval time between two continuous learning interactions. Both SAINT and SAINT+ have achieved superior performance relative to SAKT on the EdNet dataset \cite{choi2020ednet}, one of the largest publicly available datasets for educational data mining.
\newline
Similarly, \cite{ghosh2020context} observed that SAKT did not outperform DKT in their experiments and proposed a new architecture. Unlike SAINT and SAINT+, these authors presented a contextaware attentive knowledge tracing (AKT) model, incorporating self-attention mechanism with cognitive and psychometric models; these two components explain our top performance when using AKT.

\section{Conclusion}
Our main findings were the confirmation that AKT performs best on a new KT datasets compared to DKT. It would be interesting to continue studying the performance of AKT with more hyperparameter optimization to compare this baseline to the baselines previously presented.
\newline
However, the main points of improvement would likely come from changing AKT's general architecture to be able to accept more information in its design. Specifically, including timestamps information to the already exponential forgetting model would be most likely very useful input as learners tend to be very impacted by the proximity of exercises when attempting to solve them. Furthermore, the contextual information contained in questions.json like the concept routes and text could be mined to draw richer connection between concepts and enable automatic matching of concepts that might not be captured directly by the attention mechanism.

\bibliography{bibliography}
\bibliographystyle{aaai}

\end{document}